**The neurobiology of self-generated thought from cells to systems: Integrating evidence from lesion studies, human intracranial electrophysiology, neurochemistry, and neuroendocrinology**

Kieran C. R. Fox[a], Jessica R. Andrews-Hanna[b], and Kalina Christoff[a,c]

[a] Department of Psychology, University of British Columbia, 2136 West Mall, Vancouver, B.C., V6T 1Z4 Canada

[b] Institute of Cognitive Science, University of Colorado Boulder, UCB 594, Boulder, CO, 80309-0594 U.S.A.

[c] Centre for Brain Health, University of British Columbia, 2215 Wesbrook Mall, Vancouver, B.C., V6T 2B5 Canada

*Corresponding author:* Fox, K.C.R. (kfox@psych.ubc.ca)

## Abstract

Investigation of the neural basis of self-generated thought is moving beyond a simple identification with default network activation toward a more comprehensive view recognizing the role of the frontoparietal control network and other areas. A major task ahead is to unravel the functional roles and temporal dynamics of the widely distributed brain regions recruited during self-generated thought. We argue that various other neuroscientific methods – including lesion studies, human intracranial electrophysiology, and manipulation of neurochemistry – have much to contribute to this project. These diverse data have yet to be synthesized with the growing understanding of self-generated thought gained from neuroimaging, however. Here, we highlight several areas of ongoing inquiry and illustrate how evidence from other methodologies corroborates, complements, and clarifies findings from functional neuroimaging. Each methodology has particular strengths: functional neuroimaging reveals much about the variety of brain areas and networks reliably recruited. Lesion studies point to regions critical to generating and consciously experiencing self-generated thought. Human intracranial electrophysiology illuminates how and where in the brain thought is generated and where this activity subsequently spreads. Finally, measurement and manipulation of neurotransmitter and hormone levels can clarify what kind of neurochemical milieu drives or facilitates self-generated cognition. Integrating evidence from multiple complementary modalities will be a critical step on the way to improving our understanding of the neurobiology of functional and dysfunctional forms of self-generated thought.

**Introduction: Investigating the wandering brain**

One of the most intriguing yet least understood aspects of the human mind is its tendency toward ceaseless activity – a quality famously described by William James as the 'stream of consciousness' (James, 1892). This tendency of the mind to drift from one thought to another has recently sparked interest among cognitive neuroscientists and led to a growing body of neuroscientific investigations of mind-wandering, stimulus-independent thought, daydreaming, and task-unrelated thought (Mason et al., 2007, Christoff et al., 2009, Andrews-Hanna et al., 2010, Christoff, 2012, Axelrod et al., 2015). This interest is well warranted, given that these kinds of thought appear to account for as much as 30-50% of our waking thinking (Kane et al., 2007, Killingsworth and Gilbert, 2010). These various forms of undirected cognition represent a subset of a broader collection of processes referred to as "self-generated thought," defined as "mental contents that are not derived directly from immediate perceptual input" (Smallwood & Schooler, 2015; Smallwood, 2013).  Self-generated thoughts can arise spontaneously or deliberately, and their contents can be task-related or task-unrelated, as long as they arise relatively independently of immediate perceptual inputs. In this respect, emotions, mental imagery, and arguably even interoceptive signals from within the body (e.g., sensations from the stomach) can also be considered self-generated.

From the first-person perspective, self-generated thought involves a staggering variety of phenomenological content, including memory recall, future planning, mentalizing, simulation of hypothetical scenarios, and a wide variety of emotions and imagery from various sensory modalities (reviewed in Andrews-Hanna, 2012; Fox et al., 2013, 2014; Klinger, 2008; Smallwood & Schooler, 2015). Self-generated thought extends well beyond mind-wandering and daydreaming, however: self-generated mental activity is intimately involved in artistic (Ellamil et al., 2012) and scientific creativity (Maquet and Ruby, 2004), insight problem-solving (Kounios and Beeman, 2014), and dreaming (Fox et al., 2013, Domhoff and Fox, 2015). Self-generated thought is also relevant to numerous clinical, neurological, and psychiatric conditions in which typical patterns of thought are altered or exaggerated (Andrews-Hanna et al., 2014), such as depressive rumination (DuPre and Spreng, in press), Alzheimer's disease and dementia (Irish et al., 2012), post-traumatic stress disorder (Ehlers et al., 2004), and attention deficit/hyperactivity disorder (Shaw and Giambra, 1993).

Developing a comprehensive understanding of the neurobiology of self-generated thought is therefore of relevance to many fields of inquiry, from psychology to psychiatry. Early cognitive neuroscience research recognized and emphasized the importance of the default mode network (DMN) to self-generated thought (Gusnard et al., 2001, Raichle et al., 2001), but this earlier viewpoint is now giving way to a broader but also more nuanced understanding. Our recent quantitative meta-analytic treatment of the neural basis of self-generated thought, for instance, revealed no fewer than a dozen regions that appear to be consistently involved, both within and beyond the DMN (Fox et al., 2015). The most salient activations were found in the default network (including medial and rostromedial prefrontal cortex, posterior cingulate cortex, left ventrolateral prefrontal cortex, and inferior parietal lobule; Fig. 1), which has long been hypothesized to be critical to self-generated cognition in resting states (Gusnard et al., 2001, Raichle et al., 2001). Consistent recruitment was also observed, however, in frontoparietal control network regions (including dorsal anterior cingulate cortex, right anterior inferior parietal lobule, and a cluster bordering right rostrolateral and ventrolateral prefrontal cortices; Fig. 1). There were further activations that fell beyond either network, including in secondary somatosensory cortices, the left insula, medial occipital cortex (lingual gyrus), temporopolar cortex, and medial temporal lobe (Fig. 1). The inherently correlational nature of fMRI data, however, as well as its relatively poor temporal resolution, make it difficult to answer deeper questions about which of these brain regions is causally involved in generating thought, or how the origin and subsequent spread of self-generated thought appear at fine timescales on the order of milliseconds (Fox et al., 2015).

In this review we aim to synthesize a diverse body of evidence that can help begin to make sense of the role(s) played by the widely distributed regions identified by functional neuroimaging as important for self-generated thought. A key theme is that different neuroscientific modalities and methods can contribute to this project in unique but complementary ways (Table 1). Investigation of any higher cognitive process necessarily entails certain challenges, but the subjectivity and unpredictability of self-generated thought exacerbates the difficulties of conducting rigorous, well-controlled research, and underscores the importance of using multiple methodologies that can compensate for each other's limitations. Functional neuroimaging has provided an invaluable inroad into the field of self-generated cognition, but understanding the interrelationships and varied roles of the many brain areas implicated in self-generated thought requires a synthesis of evidence from many methodologies. Here, we specifically highlight contributions that can be (and already have been) made by neuropsychological lesion studies, human intracranial electrophysiology, experimental manipulations of neurotransmitter levels, and examination of hormone and other biomolecule levels. In addition to a synthesis of existing empirical data, we aim to propose some promising avenues for future research, including combining multiple methods simultaneously or in other complementary ways.

*How is self-generated thought measured?*

Experimentally, a key challenge to investigating the neural correlates of self-generated thought is knowing precisely what kind of self-generated thought is taking place, and when in time those thoughts occur. Since there is as yet no reliable objective indicator of attention being absorbed in internally-generated channels of information, cognitive neuroscientists have employed a variety of methods to investigate the content of self-generated thought – all of which necessarily rely on first-person experience reports (Fox et al., 2015). Several of the most common approaches are outlined here.

*Trait questionnaires*, such as the Imaginal Process Inventory (Singer and Antrobus, 1970), can putatively measure participants' stable (i.e., trait) self-generated thought tendencies, and have been correlated with blood-oxygenation level dependent (BOLD) activation in fMRI studies (Mason et al., 2007). *Retrospective questionnaires* typically ask subjects to characterize the average content or frequency of thought during a preceding period (Andrews-Hanna et al., 2010). *Inferential* assessment involves examining self-reported thought during various tasks or conditions completed *outside* the brain scanner. These reports are then assumed to hold for a separate sample of subjects completing the same tasks inside the scanner (e.g., Mason et al., 2007). *Online* reports involve experience-sampling probes that interrogate subjects as to their thought content in real-time (Smallwood and Schooler, 2006, Christoff et al., 2009, Schooler et al., 2011, Ellamil et al., 2016). Online reports can track the moment-to-moment content of thought in a laboratory/scanner setting (Christoff et al., 2009) or in everyday life using technologies such as smartphone applications (Killingsworth and Gilbert, 2010). This method appears to us to be the least problematic and to yield the richest data: on the one hand, it avoids the possibility of memory or self-serving biases in retrospective and questionnaire reports, respectively – and on the other hand, it can yield a large number of distinctive thought reports that can potentially reveal correspondingly distinctive neural correlates (Ellamil et al., 2016).

Despite the obvious differences between these many methods, impressive convergence has been noted, at least along several basic dimensions of self-generated thought. For instance, data gathered from questionnaires (Diaz et al., 2013), from experience sampling in everyday life (Klinger and Cox, 1987), from thoughts reported in the fMRI scanner environment (Ellamil et al., 2016), and from retrospective assessments in laboratory settings (Gorgolewski et al., 2014) all agree that visual imagery is a very prevalent component of self-generated thought. Similarly, using a wide variety of rating scales and measurement methods, nearly a dozen studies agree that emotion is prevalent in self-generated thought and that it has a mild positivity bias (Fox et al., 2014, Fox et al., in preparation).

Importantly, first steps have been taken to extend these findings to non-Western cultures, with encouraging results. For instance, a recent study examining the content of mind-wandering in the daily lives of Chinese participants (Song and Wang, 2012) found ratings of affect to be similar to those reported in numerous studies of European and North American participants (Fox et al., 2014, Fox et al., in preparation). The temporal orientation of thoughts (i.e., the proportion of thoughts about the past, present, and future, or that are 'non-temporal') was also very similar to rates reported in an American sample (Andrews-Hanna et al., 2010), and numerous studies have reported a bias toward thoughts about the future in participants from the United States (Baird et al., 2011), the United Kingdom (Smallwood et al., 2009), Belgium (Stawarczyk et al., 2011), Germany (Ruby et al., 2013), and Japan (Iijima and Tanno, 2012). All of the aforementioned results suggest that across cultures, contexts, and questionnaire methods, thought content can be relatively reliably reported by participants and assessed by experimenters.

=INSERT TABLE 1 ABOUT HERE=

=INSERT FIGURE 1 ABOUT HERE=

## Identifying the necessary and sufficient neural basis of self-generated thought: The value of neuropsychological lesion studies

'Neuropsychology' studies the cognitive, affective, and perceptual deficits suffered by patients with various brain lesions. Far from merely cataloguing the effects of rare brain disorders and diseases, however, lesion studies have shed much light on healthy brain functioning, as well as serving as highly effective catalysts for further research (e.g., (Scoville and Milner, 1957, Luria, 1976, Solms, 1997, Gainotti, 2000, Müller and Knight, 2006). Lesion studies can provide important clues about the necessary and sufficient neural basis of a given cognitive process, even a complex one like self-generated thought (Koenigs et al., 2007). To anticipate our conclusions, lesion work suggests that at least four regions play critical roles in various kinds of self-generated thought: medial prefrontal cortex, inferior parietal lobule, medial occipitotemporal cortex, and medial temporal lobe (Fig. 4).

Although we are aware of little neuropsychological work that has directly addressed the effect of brain lesions on self-generated thought, many studies have addressed closely related cognitive processes. Likely the closest parallel is the extensive neuropsychological lesion work undertaken by Solms regarding the necessary and sufficient neural basis of dreaming (Solms, 1997, 2000b). Solms concluded that the areas most critical to dreaming in general are (i) the medial prefrontal cortex and (ii) the temporoparietal junction/inferior parietal lobule. Additionally, a large swathe of (iii) medial occipital cortex, centering on the lingual gyrus, is critical for the visual features of dreaming (Solms, 1997, 2000b, Bischof and Bassetti, 2004). Elsewhere, we have argued at length that the subjective experiences and neurophysiological correlates of dreaming bear a strong resemblance to those of waking mind-wandering and related forms of self-generated thought (Fox et al., 2013, Fox and Christoff, 2014, Domhoff and Fox, 2015, Christoff et al., in press). Notably, all three regions identified by Solms as critical to nighttime dreaming emerged as significantly activated in our meta-analysis of waking self-generated thought (Fox et al., 2015; for a direct comparison, see Fig. 2, and for further discussion see Domhoff & Fox, 2015). Further supporting this view is the finding that, concurrent with the global loss or reduced frequency of dreaming, patients often reported reduced daydreaming and fantasy following the lesions to medial prefrontal cortex and/or the periventricular white matter tracts at the anterior horns of the lateral ventricles and the genu of the corpus callosum (Frank, 1946, 1950, Piehler, 1950, Schindler, 1953). To our knowledge, these latter few investigations are the only studies to have directly assessed some form of waking self-generated thought in relation to brain lesions, and unfortunately the results are largely anecdotal. Nevertheless, overall this convergence across methods (neuroimaging and lesion work; Fig. 2) suggests that more rigorous investigations of the quality and content of waking self-generated thought in patients with damage to these three areas could prove informative.

Because self-generated thought so often involves spontaneous memory retrieval (Andrews-Hanna et al., 2010, Fox et al., 2013), lesion work related to semantic and autobiographical memory capacity is also relevant. Solms' conclusions concerning the importance of the medial prefrontal cortex to self-generated waking and dreaming experience are bolstered by work showing that lesions to medial prefrontal cortex have detrimental effects on both semantic and episodic autobiographical memory retrieval (Philippi et al., 2014). Relatedly, it has recently been shown that patients with lesions in the area of the inferior parietal lobule (BA 39) report freely recalled (i.e., self-generated) memories that are impoverished and lacking in detail, despite exhibiting normal memory when probed with specific questions by the experimenters (Berryhill et al., 2007).

=INSERT FIGURE 2 ABOUT HERE=

Another potentially critical region, largely disregarded by Solms (1997), is (iv) the medial temporal lobe (Domhoff and Fox, 2015). Bilateral medial temporal lobe lesions are most famously associated with anterograde (and limited retrograde) amnesia (Scoville and Milner, 1957, Milner et al., 1968); less well known are heavy deficits in dreaming that cannot simply be explained by failures of recall. For instance, Korsakoff's syndrome patients with bilateral medial temporal lobe damage show a marked decrease in dream reports, even when awakened directly from REM sleep in a laboratory setting (Greenberg et al., 1968). Amnesia alone cannot explain these findings, for at least two reasons. First, medial temporal lobe patients still had *some* dream recall, demonstrating that they still possessed basic recall capacities (e.g., in one study, ~25% of awakenings elicited dream reports (Greenberg et al., 1968), vs. about 80-90% in normal subjects (Hobson et al., 2000). Second, medial temporal lobe patients retain intact working/short-term memory (Milner et al., 1968): anterograde amnesia should not prevent them from reporting upon experiences from just moments ago, immediately upon being awakened in a sleep laboratory setting. The dreams medial temporal lobe patients *do* report also support this interpretation: "the content of that material which *was* recalled showed very stereotyped, commonplace features and reflected very little affect" (Greenberg et al., 1968, p. 205; emphasis added). Nor are these findings restricted to Korsakoff's patients: a study of encephalitis patients with severe medial temporal lobe lesions reports identical findings (Torda, 1969): patients reported far fewer dreams than controls, and reports "were short and simple... [they] contained one scene with recurrent repetition... The dreams lacked imaginative, unusual, or mysterious details and intensive emotions. The content was stereotyped, repetitious..." (p. 280). The reduction in dream frequency with medial temporal lobe lesions is certainly severe enough to warrant further research with respect to the effect of similar lesions on waking self-generated thought.

Several studies from recent years provide further corroborative evidence for the importance of the medial temporal lobe in self-generated thought processes. For instance, medial temporal lobe patients are severely impaired in imagining novel fictitious or future events and experiences (Klein et al., 2002, Hassabis et al., 2007, Rosenbaum et al., 2009, Andelman et al., 2010, Kwan et al., 2010, Race et al., 2011, 2013), consistent with the activation of medial temporal lobe during such tasks in healthy subjects (Schacter et al., 2007, Addis et al., 2009, Schacter et al., 2012). Medial temporal lobe lesion patients also suffer deficits in tasks requiring creative and novel patterns of thinking (Duff et al., 2013, Rubin et al., 2014). Although providing only circumstantial support, these findings are intriguing because so much of the content of self-generated thought consists of imagined future scenarios and novel, creative recombination of prior memories into new thoughts and simulated experiences (Klinger, 2008, Andrews-Hanna et al., 2010, Fox et al., 2013).

Investigation of the quality and frequency of self-generated thought in medial temporal lobe lesion patients therefore appears to be a promising avenue for future research.

Major limitations of the neuropsychological method should be kept in mind, however. Lesion studies typically involve patients with large and extremely heterogeneous lesions owing to a variety of causes (e.g., stroke, traumatic brain injury, etc.), making the interpretation of such results difficult. Moreover, the simple 'overlap' method used to determine consistent lesion sites across patients, employed for instance by Solms (1997), obviously also has its limitations. In addition, brain damage typically results in some amount of recovery in the affected region, as well as structural and functional plasticity and remodeling in other brain areas in order to compensate for the deficits incurred (Johansson and Grabowski, 1994, Robertson and Murre, 1999, Kleim and Jones, 2008). This post-injury plasticity, at both the lesion site and distal brain regions, further complicates interpretations of deficits in lesion patients.

A recently-developed complement to classic lesion studies is transcranial magnetic stimulation (TMS), which allows for reversible 'lesions' of circumscribed brain areas by using pulsed magnetic fields to transiently inhibit neuronal activity (Hallett, 2000). Because TMS can in principle 'lesion' any cortical area and can be used safely in healthy people, it overcomes many of the major limitations of classic lesion studies. Although we are aware of no research that has directly investigated self-generated thought with TMS, some related work is pertinent. For example, one recent study selectively impaired self-related memory retrieval by applying TMS pulses to medial parietal default network areas (Lou et al., 2004). Related work has shown that TMS pulses to the inferior parietal lobule bilaterally, but not to medial prefrontal cortex, disrupt the classic self-reference effect (Lou et al., 2010).

Conversely, the process of self-*evaluation* has been shown to be selectively disrupted by TMS applied to medial prefrontal cortex (Luber et al., 2012) – potentially suggesting a double-dissociation in terms of distinct aspects of self-related processing across medial prefrontal and parietal areas. While of course preliminary, these results point to the potential power of TMS to directly target a variety of regions implicated in self-generated thought and systematically address their specific functional contributions.

**Dynamics of self-generated thought: Neural origins and ontogeny as revealed by human intracranial electrophysiology**

Neither functional neuroimaging nor lesion studies can answer questions about the detailed, millisecond-scale temporal dynamics of self-generated thought. Two key questions about these temporal dynamics concern the neuroanatomical *origins* and *ontogeny* of self-generated thought: where in the brain do self-generated thoughts tend to originate (there may of course be more than one answer), and how does self-generated activity subsequently spread through distributed neuronal networks and give rise to the accompanying subjective experiences of memory recall and novel thought? A third important question centers on the dynamics of interactions between large-scale networks, most importantly the frontoparietal control and default mode networks, which we discuss elsewhere (Christoff et al., in press).

Human intracranial electrophysiology or electroencephalography (iEEG) offers an ideal means to investigate questions of neural origins and ontogeny. Typically, iEEG is implemented for the assessment, diagnosis, and treatment of otherwise intractable neurological conditions, such as epilepsy or Parkinson's disease (Bechtereva and Abdullaev, 2000, Lachaux et al., 2003). Most of the data relevant to the present review come from the study of epilepsy patients: in an effort to identify the precise epileptogenic focus in brain tissue, numerous microelectrodes can be chronically implanted (for weeks or more) in various brain regions thought to be involved in seizure generation (Fried et al., 2014). As the medial temporal lobe is a common origin site for epileptic seizures, microelectrodes are often implanted into deep cortical and subcortical areas, including the amygdala, hippocampus, and entorhinal cortex (Vignal et al., 2007, Gelbard-Sagiv et al., 2008). This technique allows direct recording of various aspects of the brain's neuroelectric activity, most commonly measurement of the local field potential (i.e., extracellularly-recorded electrical potential fluctuations) thought to largely reflect summation of nearby synaptic currents in dendrites and neuron somata, and to a lesser extent axon potential firing (Elul, 1972, Buzsáki et al., 2012). Advances in microelectrode technology and analysis methods have also allowed for the reconstruction of putative single-neuron spiking activity from extracellularly recorded potentials (Gelbard-Sagiv et al., 2008, Fried et al., 2014). When we refer to brain 'activity' throughout this section, we therefore mean either single-neuron spiking or changes in the local field potential, as measured with microelectrodes placed in the extracellular matrix.

A related method involves mild electrical stimulation of the cortical surface with (non-implanted) electrodes during neurosurgical operations, typically undertaken in an effort to identify and spare cortical areas critical to motor output generally and speech in particular (Penfield and Boldrey, 1937, Penfield and Welch, 1951, Penfield, 1958). Alternatively, a larger grid or strip of electrodes can be placed on the cortical surface for the same purpose, known as electrocorticography or 'eCoG' (Crone et al., 2006, Buzsáki et al., 2012).

Because patients are awake and alert with both chronic implanted microelectrodes and during epilepsy surgeries (typically conducted under only local anesthesia), first-person experience reports can be collected and correlated with the location either of stimulation or passively-recorded discharges and increases in spiking activity. The correlation of brain activity with subjective experience can be undertaken in any of several ways. In some studies, patients can simply describe their experiences in an open-ended way as stimulation is administered to particular locations (Penfield and Perot, 1963, Parvizi et al., 2013); in other cases, spontaneous subjective experiences (e.g., free memory recall or epileptic aura phenomena) can be indicated by the patient, and self-generated brain activity can then be analyzed in the time window surrounding these subjective reports (Vignal et al., 2007, Gelbard-Sagiv et al., 2008). Although the number of electrodes and experiential reports is generally small for any given patient, general principles can often be gleaned by synthesizing data from large numbers of such investigations, involving widespread electrode placement in many individuals (Penfield and Perot, 1963, Selimbeyoglu and Parvizi, 2010, Burke et al., 2014). Here we compare and synthesize the results of dozens of independent studies in an effort to delineate whether they provide clues to the neural origins and subsequent ontogeny of self-generated thought.

*Neural origins of self-generated thought*

With respect to sites of origin, there may be more than one central location of generation, or it may be that difficult-to-localize, distributed network activity gives rise to thoughts, and the search for 'primary' thought generation regions is misguided. The reality could also be that both mechanisms contribute, depending on the type and content of self-generated thought. Whatever the answers to these quandaries, the evidence for the central role of the medial temporal lobe in thought generation is compelling. The most direct causal evidence comes from the ever-growing body of cognitive studies of patients who have had brain electrodes chronically implanted, or the cortical surface probed with electrical stimulation, for a variety of clinical reasons – usually intractable epilepsy. Since the first studies of this kind nearly one hundred years ago, essentially every brain area has been explored with direct focal electrode stimulation, or passive recording of spontaneous discharges, to a greater or lesser degree. The most relevant finding from this growing body of research is that stimulation of medial temporal lobe structures (i.e., the hippocampus, parahippocampus, entorhinal cortex, and amygdala) very frequently leads to memory recall, immersive thoughts, and hallucinatory, dream-like experiences (Fig. 3). Far from an isolated occurrence, the accumulated evidence supporting this assertion is fairly substantial (Feindel and Penfield, 1954, Bickford et al., 1958, Baldwin, 1960, Penfield and Perot, 1963, Horowitz et al., 1968, Ferguson et al., 1969, Halgren et al., 1978, Wieser, 1979, Gloor et al., 1982, Fish et al., 1993, Bancaud et al., 1994, Kahane et al., 2003, Barbeau et al., 2005, Vignal et al., 2007, Mulak et al., 2008, Jacobs et al., 2012).

Equally important, however, is the specificity of these results: a comprehensive review of cognitive-affective findings from electrophysiological studies in humans, summarizing the results of over 100 such studies conducted over the past eighty years, found that such phenomena were produced almost exclusively by stimulation of medial temporal lobe structures (Selimbeyoglu & Parvizi, 2010; see our summary of relevant results in our Table 2). Stimulation of nearby temporopolar cortex (Penfield and Perot, 1963, Halgren et al., 1978, Bancaud et al., 1994) or lateral temporal cortex (Penfield, 1958, Mullan and Penfield, 1959, Penfield and Perot, 1963, Bancaud et al., 1994) can also elicit such phenomena, but these reports are comparatively rare, and moreover electrical stimulation is known to spread to adjacent cortical areas, making it difficult to strictly rule out a medial temporal origin (or co-activation) in many of these cases (Gloor et al., 1982, Gloor, 1990, Bancaud et al., 1994). Indeed, many of the stimulations to lateral temporal cortex that elicited thought- and dream-like experiences were reported from electrodes at a depth of several centimeters, supporting such a possibility (Selimbeyoglu and Parvizi, 2010). Stimulation of lateral prefrontal cortex (Blanke et al., 2000a) and orbitofrontal cortex (Mahl et al., 1964) can likewise occasionally elicit such experiences, but such reports are few and far between (Table 2). Most striking is that no such experiences appear to have ever been reported from stimulation of virtually any other area in the brain (Selimbeyoglu and Parvizi, 2010), even from the many other regions consistently recruited by self-generated thought in functional neuroimaging investigations (Fig. 3). For instance, fewer than 1% of stimulations to the inferior parietal lobule elicit such phenomena (Selimbeyoglu and Parvizi, 2010). We summarize these results in Table 2 and Fig. 3.

Although absence of evidence is not necessarily evidence of absence, the results are intriguing given the number of investigations already carried out and the span of time over which such investigations have been taking place (well over a hundred studies, over nearly a hundred years). Together these results strongly suggest that the medial temporal lobe is a key generation site for many forms of self-generated thought, including dreaming – a hypothesis that could be examined more forcefully with lesion patients (see previous section).

A central limitation of drawing conclusions based on this body of research is that the results were primarily *evoked* by electrical stimulation – they mostly do not represent self-generated, spontaneous brain activity giving rise to immersive thoughts and hallucinatory, dream-like experiences. Several studies, however, *have* passively recorded spontaneous brain activity in conjunction with first-person reports of accompanying experience, and yielded consistent findings: the medial temporal lobe appears to be by far the most common origin of self-generated brain activity accompanied by spontaneous memories, thoughts, and dream-like experiences (Bancaud et al., 1994, Vignal et al., 2007, Gelbard-Sagiv et al., 2008).

=INSERT TABLE 2 AND FIGURE 3 ABOUT HERE=

*The subsequent ontogeny of self-generated thought*

Even highly localized stimulation or spontaneous neural firing will tend to spread to proximate and distant brain areas through short- and long-range connections (Halgren and Chauvel, 1992, Selimbeyoglu and Parvizi, 2010). Assuming that the medial temporal lobe is one of the main sources of spontaneous neuronal activity giving rise to thoughts and spontaneously recalled memories, the question remains of how and on what timescale this activity spreads, and where in the brain it is most likely to spread *to*. Given that the medial temporal lobe is well connected to nearly every other part of the brain (Buzsáki, 1989, Rolls, 2000, Simons and Spiers, 2003, Buzsaki, 2006), the possibilities are legion.

Our view of the role of the medial temporal lobe in spontaneously reactivating memory traces, and in recombining mnemic material into novel thoughts and imaginings, draws heavily on the hippocampal indexing theory (Teyler and DiScenna, 1985, 1986, Moscovitch, 1992, Teyler and Rudy, 2007). In essence, "indexing theory proposes that the content of our experiences is stored in the multiple neocortical loci activated by experience and the hippocampus stores an index of those neocortical loci" (Teyler & Rudy, 2007; p. 1160). Indexing theory is primarily concerned with accounting for the reactivation of *memories*; our view expands on this to account also for novel patterns of thought that draw on mnemic material but recombine it into new thoughts and simulations of experience. If a specific pattern of medial temporal lobe activity and connectivity to neocortex indeed gives rise to memory recall, it follows that slightly altered or novel patterns of hippocampal activity will 'index' novel patterns of distributed neocortical activity never before experienced – that is, novel thoughts and simulated experiences. Although this account is speculative, it is consistent with lesion evidence (discussed above) indicating that MTL is crucial for the detailed imagination of novel or fictitious scenarios (Klein et al., 2002, Hassabis et al., 2007, Rosenbaum et al., 2009, Andelman et al., 2010, Kwan et al., 2010, Race et al., 2011, 2013). Because dreams almost never replay actual episodic memories (Fosse et al., 2003), but rather recombine mnemonic elements into novel scenes and experiences (Nielsen and Stenstrom, 2005), the lesion evidence that MTL is crucial to dream generation (Greenberg et al., 1968, Torda, 1969) also supports its putative role in generating novel mental content. Moreover, it has been suggested that the patterns of synaptic connectivity within the hippocampus in particular lend themselves well to novel connections. In contrast to the connectivity of most other cortical areas, where short-distance synapses to nearby cells predominate and longer-distance connections rare (Thomson and Bannister, 2003, Douglas and Martin, 2004, Markram et al., 2004), hippocampal neurons are almost equally likely to contact nearby and distant neighbors (Li et al., 1992, Li et al., 1994, Buzsaki, 2006). This widely divergent microcircuitry means that any neuron can contact any other with a minimal number of synapses (usually no more than 2-3) – an appealing neural substrate for arbitrary, unlikely, or novel connections between neurons which otherwise represent highly distinctive perceptual or mnemonic qualities (Buzsaki, 2006). We suggest that spontaneous activity in MTL regions can therefore potentially activate novel neuronal networks within the MTL, which can in turn 'index' and recruit novel patterns of activity throughout the brain through the MTL's dense interconnections with other areas.

The central problem is to understand "the manner in which electrical stimulation of a particular location [or spontaneous, self-generated activity at a given location] leads to activation of [a] very widespread but, at the same time, very particular network" (Bancaud et al., 1994; p. 87). We do not predict a single, universal pattern of spreading activity. The pattern of subsequent recruitment should instead be related to the experiential qualities of the accompanying thoughts, including the sensory modalities instantiated (visual, auditory, somatosensory, etc.); affective tone (positive, negative, or neutral); and other qualities such as temporal orientation (past, present, future) and goal-relatedness. For instance, thoughts that are visual in nature may involve spreading activity (at the neuronal level) and subsequently observable recruitment (e.g., with functional neuroimaging) of areas such as the lingual gyrus (cf. Fig. 1). Similarly, we would predict that more goal-related thoughts involving planning for the future, or thoughts that are otherwise guided in an intentional and voluntary fashion, should tend to recruit prefrontal executive areas such as dorsal anterior cingulate cortex and rostrolateral prefrontal cortex (Seli et al., 2016). Some preliminary evidence for the neural dissociability of different thought types has been provided by fMRI investigations: for instance, using multivariate pattern analyses (MVPA), one group was able to predict the emotional valence (positive vs. negative) of thoughts at above chance levels based on activation in the medial prefrontal cortex (Tusche et al., 2014); another study found varying patterns of intrinsic brain activity associated with a variety of thought tendencies, such as the frequency of visual or future-oriented thoughts (Gorgolewski et al., 2014). For more on this issue, see the Discussion section in Fox et al. (2015).

A further consideration is that new thoughts and imagined mental content may not be determined merely by patterns of synaptic connections alone – these anatomical connections could also interact with a host of other factors that all participate in the sculpting of spontaneous and 'noisy' brain activity into an overall pattern of activity ultimately corresponding to subjectively experienced mental content of one form or another. Aside from the basic neurochemical state of the brain at a given time (see next section), "current sensory input, cognitive context and/or psychosocial concerns could sometimes be major influences in defining the final pattern to emerge from the sculpting" (Bancaud et al., 1994; p. 87).

Evidence bearing on this question is unfortunately much more sparse – commensurate with the increased difficulty of the problem. A comprehensive understanding of the spread of spontaneous activations would require widespread electrode placement throughout the brain – a situation uncommon in human electrophysiology studies, where electrode placement is determined strictly by clinical criteria and rarely requires simultaneous recording from many widely dispersed sites.

To truly explore the hypothesized model of self-generated thought ontogeny requires recording from relevant cortical areas *and* the medial temporal lobe, simultaneously. The study that probably best approximates this ideal methodology investigated episodic memory retrieval while recording simultaneously from medial temporal lobe and the retrosplenial cortex in the posteromedial area (Foster et al., 2013). The authors reported that episodic memory retrieval involved phase locking in the theta band (3-4 Hz) between these two regions, and that this concerted activity was unique to this frequency band as well as to these two regions. Most intriguing is that the coupling was strongest *prior* to actual peak high-frequency activity in the retrosplenial cortex, potentially suggesting a primary role for the medial temporal lobe. Another study investigating spontaneous memory recall with human intracranial electrophysiology found that medial temporal lobe structures demonstrated some of the strongest increases in high-frequency (i.e., high $\gamma$–band) activity just *prior* to conscious recall and reporting of the memory (Burke et al., 2014), whereas high-frequency activity peaked *later* in several parietal, temporal, and frontal regions (see their Fig. 4). Importantly, although the medial temporal lobe was not the only region in which high-frequency activity peaked prior to recall, it was the only one in which this activity significantly predicted subsequent memory recall, suggesting a cardinal role in memory recollection (Burke et al., 2014).

A limitation of these studies (for our purposes) is that they involve memory retrieval in one form or another; therefore they do not speak to the possibility of truly *novel* patterns of activity – a precondition for any neuron-level model of self-generated thought. Some preliminary evidence for this possibility, however, comes from animal models. Several well-known animal studies have demonstrated that patterns of medial temporal lobe activity reflecting recent spatiotemporal experiences and memories are spontaneously replayed during periods of resting wakefulness (Foster and Wilson, 2006, Diba and Buzsáki, 2007) or subsequent sleep (Wilson and McNaughton, 1994). Coordinated replay across medial temporal lobe and various neocortical areas, including visual regions (Ji and Wilson, 2007) and posterior parietal cortex (Qin et al., 1997), has also been reported – although a causal role for the medial temporal lobe in initiating this activity has not been shown. Interestingly, recent studies have shown that 'replay' need not be a mere recapitulation of previous firing patterns, but often contains novel firing sequences that do not correspond to any actual spatiotemporal sequence of experience (Davidson et al., 2009, Gupta et al., 2010). These novel firing sequences have been interpreted as evidence for planning and imagining of alternative behaviors (Knierim, 2009, Gupta et al., 2010) – i.e., the generation of novel thoughts, plans, and imagined scenarios, akin to self-generated thought content in humans.

In summary, although the network-level ontogeny of thoughts putatively generated in the medial temporal lobe remains largely obscure, many important prerequisites for an 'indexing' scenario have already been demonstrated. The medial temporal lobe is known to spontaneously reactivate patterns of activity first instantiated during novel experience and learning, and this reactivation takes place not only during sleep but also waking behavior and restful states. This patterned replay can be temporally synchronized with various neocortical brain areas (Qin et al., 1997, Ji and Wilson, 2007), and, critically, 'replay' can in fact involve novel patterns of activity that do not correspond to any specific experience (Davidson et al., 2009, Knierim, 2009, Gupta et al., 2010). Although still far from definitive, all of these findings are consistent with an 'indexing' theory of self-generated thought origin and ontogeny, whereby spontaneous medial temporal lobe activity activates neuronal networks dispersed throughout the brain, instantiating either recall of memories or the experience of novel thoughts and imaginings. Perhaps most important, electrode montages that include both medial temporal lobe and other regions clearly important to self-generated thought (Fig. 1) are occasionally employed in clinical settings with human participants (Ekstrom et al., 2003, Gelbard-Sagiv et al., 2008, Foster et al., 2013, Burke et al., 2014); the stage is therefore set for further exploration of the temporal and spatial dynamics that characterize the ontogeny of thoughts in the human brain. The most fruitful approach might be to combine intracranial electrophysiology (which has unparalleled temporal resolution, but cannot be expected to cover a large number of brain regions in human studies) with functional MRI (with poor temporal resolution, but the ability to sample from all brain regions simultaneously). The safe and optimal combination of these methods is an area of active research and has already been demonstrated in feasibility studies (Carmichael et al., 2007, Carmichael et al., 2010).

*Limitations of intracranial EEG investigations in humans*

Some important limitations of the aforementioned research, and hence the conclusions drawn from it, should be kept in mind. First, all such recordings and stimulations have, by necessity, been conducted in patients with severe neurological conditions. Although in some cases these conditions might not be expected to appreciably alter neuroelectric activity in and of itself, because results cannot be compared with healthy controls the possibility remains of unknown and unforeseen differences that limit the generalizability of any findings to healthy humans. Second, the number and placement of electrodes is determined strictly by considerations of clinical necessity in any given patient. While this practice is of course indisputably the right course of action, for purposes of research on cognition it can often mean that electrodes are too few, or not optimally placed, to answer research questions. This leads to a third complication, namely drawing conclusions based on concatenation of data from large numbers of patients with disparate neurological conditions and highly variable electrode placement. Although of course this can be a powerful approach yielding novel insights (such as those we have proffered, above, on neural origins and ontogeny of self-generated thought), nonetheless it should be kept in mind that synthesis of data on this scale introduces large but not-fully-understood sources of variability that might limit or distort any putative findings.

**The modulation of self-generated thought: Indications from the neurochemistry of REM sleep and creative thinking**

All brain activity, from single-neuron spike trains to large-scale network interactions, takes place within the context of a neurochemical milieu (Brady et al., 2005). Even relatively small differences in levels of key neurotransmitters and neuromodulators can have profound effects on cognition, affect, and even consciousness itself (Perry et al., 2002). Everyday externally-oriented waking cognition is intimately dependent on a particular neurochemical profile – and alterations in absolute and relative neurotransmitter levels rapidly leads to strikingly different mental states, such as deep sleep, dreaming, anesthesia, and so on (Jones, 1991, Perry et al., 2002, Jones, 2005). Waking self-generated thought typically involves a decoupling from the external perceptual environment (Smallwood et al., 2008, Kam et al., 2011, Kam et al., 2013, Kam and Handy, 2013), suggesting deviations from the 'standard' neurochemistry that supports vigilant monitoring of, and engagement with, the outside world (Smallwood et al., 2012). Sleep and dreaming represent the limit cases of this kind of perceptual disengagement, and are driven by a neurochemical profile that differs drastically from waking (Jones, 1991, 2005). Dreaming is a particularly intriguing parallel because, like waking self-generated thought, it involves a simultaneous decoupling from external inputs alongside self-generated simulations, perceptions, and affect. Waking self-generated thought therefore represents an intriguing in-between state involving partial disengagement from external inputs alongside heightened attention to inner channels of information (Dixon et al., 2014). Together with our functional neuroimaging findings (reviewed above; Fig. 2) that waking self-generated thought shows patterns of brain activation intermediate between externally-oriented thinking and fully decoupled, internally-generated dream mentation (Fox et al., 2013, Domhoff and Fox, 2015, Fox et al., 2015), these results hint at the possibility that waking self-generated thought is characterized, and perhaps influenced to a large degree, by a unique neurochemical profile (Hu et al., 2013, Mittner et al., 2014). Findings of alterations in self-generated thought following ingestion of substances support this notion: for instance, consuming moderate levels of alcohol increases rates of probe-caught mind-wandering while simultaneously reducing meta-awareness of mind-wandering as indexed by self-reported zoning out (Sayette et al., 2009). Although the exact neurochemical mechanisms of action of alcohol in the brain remain poorly understood, there is no doubt that it has significant effects on multiple neurotransmitter systems, including GABA receptor agonism and NMDA receptor antagonism (Krystal and Tabakoff, 2002). Intriguingly, the dissociative anesthetic ketamine also appears to act primarily via antagonism of NMDA receptors (Salt et al., 1988, Jansen and Sferios, 2001) and can likewise lead to increased attention to self-generated channels of information (Fox et al., in press).

Beyond these suggestive findings, however, little research has directly considered the neurochemistry of self-generated thought. Although investigation of the neurochemical determinants and correlates of self-generated thought therefore remains a largely unexplored field of research, promising avenues of research are suggested by investigations of the neurochemistry of several related cognitive processes (Christoff et al., 2011). The first is dreaming, which (as explained above) we view as self-generated thought *par excellence* (Fox et al., 2013, Domhoff and Fox, 2015, Christoff et al., in press). Although dreaming is in principle doubly dissociable from REM sleep (Solms, 2000b), in practice dreaming almost always accompanies REM ($r = .8$), and occurs far more frequently during this sleep stage than any other (Hobson et al., 2000, Fox et al., 2013). Further, although prefrontal executive areas are deactivated during REM sleep (whereas some activation of executive areas is retained in waking self-generated thought), many of the areas active during self-generated thought are even *more* active in REM, consonant with the more immersive and hyper-associative experiences of dreaming (as compared to waking thought and fantasy). In practice, then, the neurochemistry of REM sleep is potentially informative for future investigations of the neurochemistry of waking self-generated thought, in that it may shed light on the brain states and neurotransmitter ratios that facilitate highly associative and novel patterns of thought, and increased attention to internal channels of information. Although the neurochemistry of sleep is extremely complex, and moreover much of the data is drawn from animal research due to the difficulty of studying neurotransmitter levels in humans, nonetheless some general conclusions have been proposed. The general trend in REM sleep appears to be a decrease in levels of most major neurotransmitters and neuromodulators (e.g., GABA, histamine, glutamate, serotonin, etc.) – with the marked exception of acetylcholine and dopamine, which instead appear to be elevated beyond, or at least on par with, normal waking levels (Gottesmann, 1999, Pace-Schott and Hobson, 2002, Solms, 2002, Jones, 2005, Lena et al., 2005, Hobson, 2009). Behavioral evidence from humans provides some corroboration of these largely animal-derived values: for instance, increased vividness and emotionality of dreams has been reported in patients with Alzheimer's disease or dementia being treated with the acetylcholinesterase inhibitor galantamine (Corbo et al., 2013), and recreational users of galantamine have provided similar, if anecdotal, reports (Laberge, 2004, Yuschak, 2006). There have also been reports of increased vividness, duration, and emotionality of dreaming in Parkinson's disease patients being treated with L-dopa, a precursor molecule of dopamine (as well as other neurotransmitters) (Moskovitz et al., 1978, Sharf et al., 1978, Solms, 2002). Given the many similarities between dreaming and waking self-generated thought, these neurochemical findings, although tentative, suggest hypotheses about the neurochemistry of self-generated thought that would be relatively straightforward to test. For instance, experimental administration of safe, reasonable doses of dopamine precursors, or acetylcholinesterase inhibitors, could be compared with placebo in terms of effect on the frequency, affective tone, or vividness of self-generated thought.

Complementary hypotheses can be derived from studies of the neurochemistry of creative thought. The similarities in terms of content, cognitive process, and neurophysiological recruitment between self-generated thought and creativity are discussed in detail elsewhere (Christoff et al., 2011, Ellamil et al., 2012, Fox and Christoff, 2014, Beaty et al., 2015); the salient point for us here is that these several similarities make the neurochemistry of creative thought potentially informative. Probably the most reliable finding from such research is that decreasing levels of arousal-heightening neurotransmitters, such as norepinephrine, appears to be beneficial for creative thinking (Beversdorf et al., 1999, Heilman et al., 2003, Silver et al., 2004, Christoff et al., 2011). The proposed rationale for these findings is that decreased arousal and cognitive control facilitates the associative, novel forms of thought necessary to the generation of creative ideas (Christoff et al., 2011). Given that self-generated thought commonly involves a similar creative recombination of past ideas and memories into novel simulations and imagination, a straightforward hypothesis would be that norepinephrine antagonists could, similarly, lead to increased frequency or novelty of self-generated thought.

Related to this discussion, a specific proposal has recently been put forward that self-generated thought may be facilitated by high tonic norepinephrine activity (Smallwood et al., 2012). The key distinction here is between vigorous *phasic* norepinephrine activity, which appears to facilitate attention to changing goals in the external world, and high *tonic* levels, which might reduce the signal-to-noise ratio of external stimuli and allow for greater attention to internal channels of information, such as self-generated thought (Smallwood et al., 2012). Although this hypothesis may not necessarily align perfectly with predictions based on creativity research, both lines of thinking agree on the importance of further investigations of the role played by norepinephrine.

To summarize, the unique constellation of brain networks activated by self-generated thought, together with its singular phenomenological properties midway between the external and internal worlds, suggests an uncommon neurochemical profile – but one that may have parallels in other cognitive states. Evidence from cognitive processes that exhibit patterns of brain recruitment similar to self-generated thought, including REM sleep and dreaming (Fox et al., 2013) and creative thinking (Ellamil et al., 2012, Beaty et al., 2015), suggest that investigation of acetylcholine, dopamine, and norepinephrine in particular could prove fruitful. Hypotheses about the enabling or inhibitory role of these neurotransmitters in self-generated thought could be tested relatively easily using simple thought sampling paradigms (Christoff et al., 2009) combined with the administration of safe, widely-available, and affordable neurochemical agents and precursors (cf. (Chamberlain et al., 2006).

**Systems biology of self-generated thought: Relationship to neuroendocrinology and other biomolecules throughout the body**

Although this review has focused mostly on neurobiology, clearly the brain is embedded in, and interacts intimately with, the rest of the body: what the brain does and thinks about affects, and is affected by, the broader biological state of the organism and levels of countless hormones and other biomolecules. This systems biology perspective understands the entire body, including the nervous system, as a deeply intertwined and mutually interdependent set of subsystems with complex, nonlinear, and difficult-to-predict effects on one another (Kitano, 2001, Capra and Luisi, 2014). Such a perspective can begin to shed light on mostly unexplored relationships between mental content, brain activation, and organism-wide biochemistry. Here we discuss some examples of seminal work in this domain examining neuroendocrinology and other biomolecules at the sub-cellular level.

*Neuroendocrinology of self-generated thought*

The neuroendocrinology of self-generated thought explores how various hormones may affect, or be affected by, the content and general pattern of one's thought. A straightforward prediction would be that negatively-toned self-generated thought might result in (or from) elevated levels of stress-related hormones, such as cortisol (Chrousos, 2009). One recent study explored the affective valence of self-generated thought and its relationship to cortisol and α-amylase, two biomolecules implicated in the body's basic stress response mediated by the hypothalamic-pituitary-adrenal (HPA) axis (Engert et al., 2014). The authors found that the frequency of negatively-valenced as well as past-oriented thoughts was associated with elevated levels of both biomarkers. These findings parallel earlier research that also found relationships between elevated cortisol and higher levels of rumination, a hallmark of depressive thinking (Zoccola et al., 2008, Rydstedt et al., 2009). Clearly much remains to be done toward understanding the neuroendocrinology of self-generated thought, but these pioneering studies demonstrate that such investigations are both feasible and potentially informative.

*Other biomolecules and self-generated thought*

Another relatively unexplored area of research extends beyond hormones to an endless array of additional biomolecules that might be related to the content and overall patterns of self-generated thought. One pioneering study has examined the relationship between frequency of self-generated thought and the length of telomeres, the repetitive nucleotide sequences that protect chromosome termini (Epel et al., 2012). Telomere length decreases with natural aging, but also in association with psychological and physiological stress, with telomere shortness predicting early disease onset and mortality (Lin et al., 2012). The authors found that higher frequency of self-reported mind-wandering was associated with significantly shorter immune cell telomere length, and this relationship persisted even after controlling for other sources of stress (Epel et al., 2012). These intriguing results only hint at the vast number of meaningful interrelationships that might exist between patterns of thought content and brain activation on the one hand, and biomarkers of health, stress, and aging on the other.

## Conclusion

The discovery of the DMN has been extremely influential in drawing attention to intrinsic brain activity and the self-generated mental experience that often accompanies it (Raichle, 2010). Functional neuroimaging has provided sufficient empirical evidence to give a general, if preliminary, notion of the neural basis of self-generated thought, which includes but also extends beyond the DMN (Fig. 1). A wide array of neuroscience methods is now needed to more fully understand the wide variety of brain regions and multiple networks implicated in self-generated forms of cognition, as well as the effect self-generated thought has on the rest of the body. As the nascent cognitive neuroscience of self-generated thought develops, research needs to move beyond DMN-based models to begin investigating its many other neurobiological characteristics with methods such as lesion studies (Fig. 2), intracranial electrophysiology (Fig. 3), and experimental manipulation and measurement of neurotransmitters and relevant hormones.

The spectacular success of the DMN has served as a ladder of sorts by which cognitive neuroscience has reached a much wider acceptance of the value of studying spontaneous brain activity and its relationship to self-generated thinking. From this higher vantage point, the DMN can now be seen as a critical, but only partial, component of the neurobiology of self-generated thought. A DMN-centric view has helped us climb to this broader perspective, but may be in danger of becoming an obstacle to further progress. To paraphrase Wittgenstein: To climb beyond we must, so to speak, throw away the ladder after we have climbed up it (Wittgenstein, 1994).

**Acknowledgements**

This work was supported by a Natural Sciences and Engineering Research Council (NSERC) Vanier Canada Graduate Scholarship awarded to K.C.R.F.; by grants from the National Institutes of Mental Health (1F32MH093985), the Brain & Behavioral Research Foundation, and the Templeton Foundation (Science of Prospection Award) awarded to J.R.A-H.; and grants from the Canadian Institutes of Health Research (CIHR) and NSERC awarded to K.C. The authors declare no conflicts of interest.

*Table 1.* Major areas of inquiry in the neurobiology of self-generated thought.

| Aspect of Self-Generated Thought | Central Questions | Principal Modality of Investigation |
|---|---|---|
| Breadth and Diversity | What brain regions and networks are consistently recruited by self-generated thought? How does the diversity of recruitment vary with the content and type of self-generated thought? | Functional neuroimaging (fMRI and PET) |
| Necessary Substrate | What regions of the brain form the necessary (if not sufficient) neural substrate allowing for the self-generation of thought? | Neuropsychological lesion studies; transcranial magnetic stimulation |
| Origin and Ontogeny | How and where is thought self-generated in the brain? How do initial patterns of activity recruit and interact with other parts of the brain? | Human intracranial electrophysiology; magnetoencephalography |
| Modulation | What neurochemical milieu drives or facilitates self-generated thought? Does the neurochemistry of self-generated thought resemble that of kindred cognitive processes such as dreaming or creative thinking? | Manipulation and measurement of neurotransmitter and neuromodulator levels |
| Systems Biology | What other biomolecules beyond the central nervous system affect or are affected by self-generated thought? Are there relationships between self-generated thought content and neuroendocrinology or other biomarkers throughout the body? | Manipulation and measurement of hormones and other biomolecules (e.g., enzymes) |

*Table 2.* Summary of human electrophysiology studies demonstrating elicitation of memories, thoughts, or hallucinatory, dream-like experiences.

| Brain Region | Stimulations/ discharges eliciting[a] | Total stimulations/ discharges | Percentage eliciting | References |
|---|---|---|---|---|
| **Temporal Lobe** | | | | |
| Hippocampus | 25 | 46 | 54% | (Halgren et al., 1978, Fish et al., 1993, Bancaud et al., 1994, Kahane et al., 2003, Vignal et al., 2007, Mulak et al., 2008) |
| Amygdala | 13 | 36 | 36% | (Ferguson et al., 1969, Halgren et al., 1978, Fish et al., 1993, Vignal et al., 2007) |
| Parahippocampal region | 9 | 16 | 56% | (Feindel and Penfield, 1954, Penfield and Perot, 1963, Vignal et al., 2007) |
| Temporopolar cortex | 5 | 11 | 45% | (Penfield and Perot, 1963, Halgren et al., 1978, Bancaud et al., 1994, Ostrowsky et al., 2002, Mulak et al., 2008) |
| Inferior temporal gyrus | 1 | 21 | 5% | (Penfield and Perot, 1963) |
| Middle temporal gyrus | 7 | 42 | 17% | (Penfield, 1958, Mullan and Penfield, 1959, Penfield and Perot, 1963, Kahane et al., 2003) |
| Superior temporal gyrus | 24 | 99 | 24% | (Mullan and Penfield, 1959, Penfield and Perot, 1963, Morris et al., 1984) |
| Temporo-occipital junction | 4 | 17 | 24% | (Penfield and Perot, 1963, Morris et al., 1984, Lee et al., 2000) |
| **Frontal Lobe** | | | | |
| Inferior frontal gyrus | 1 | 7 | 14% | (Blanke et al., 2000a) |
| Middle frontal gyrus | 2 | 8 | 25% | (Blanke et al., 2000a) |
| Orbitofrontal cortex | 1 | 4 | 25% | (Mahl et al., 1964) |
| Supplementary motor area | 1 | 6 | 17% | (Beauvais et al., 2005) |
| **Parietal Lobe** | | | | |
| Inferior parietal lobule | 2 | 42 | .05% | (Blanke et al., 2000b, Schulz et al., 2007) |

Based on data in Supplementary Table 1 in the comprehensive review conducted by Selimbeyoglu & Parvizi (2010). Data for brain areas with ≥10 stimulations/discharges reported in the literature are visualized in Fig. 3.

*Figure 1.* The breadth of self-generated thought recruitment (green clusters) juxtaposed with the default mode network (blue borders) and frontoparietal control network (red borders).

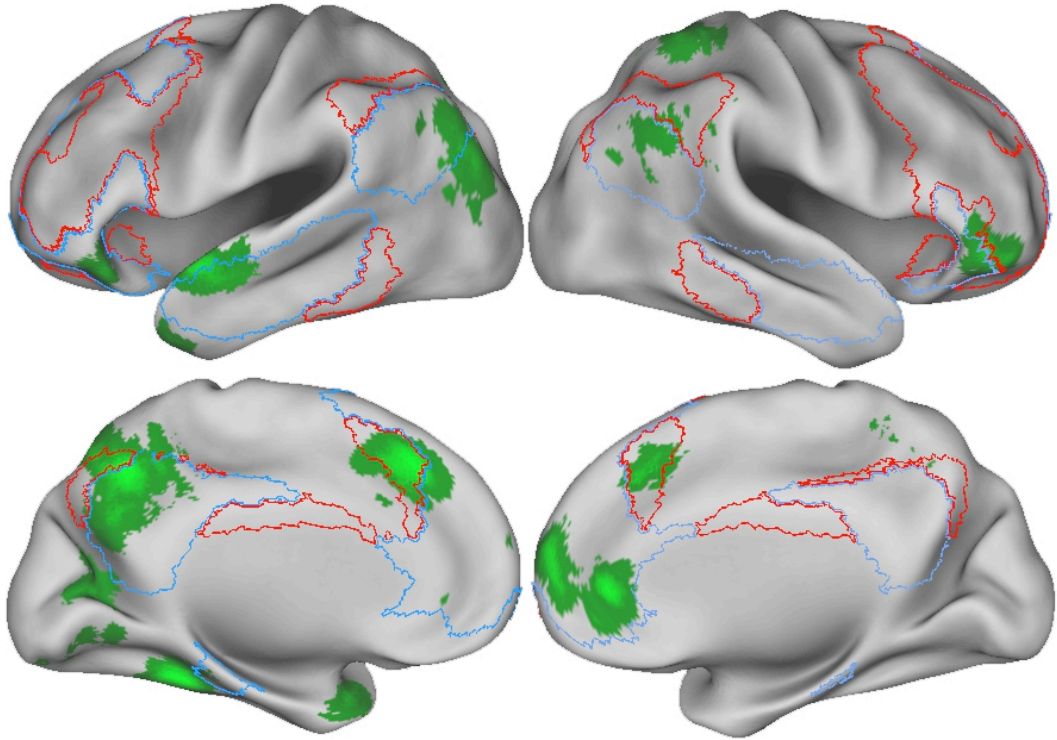

Cortical mapping of significant meta-analytic clusters associated with mind-wandering and related self-generated thought processes (green clusters) juxtaposed with outlines of the default mode network (blue) and the frontoparietal control network (red). Note that self-generated thought activations overlap considerably with both networks, but also include regions beyond both networks (highlighted in Figure 1). Default mode network and frontoparietal control network masks based on Yeo et al. (2011). Reproduced with permission from Fox et al. (2015).

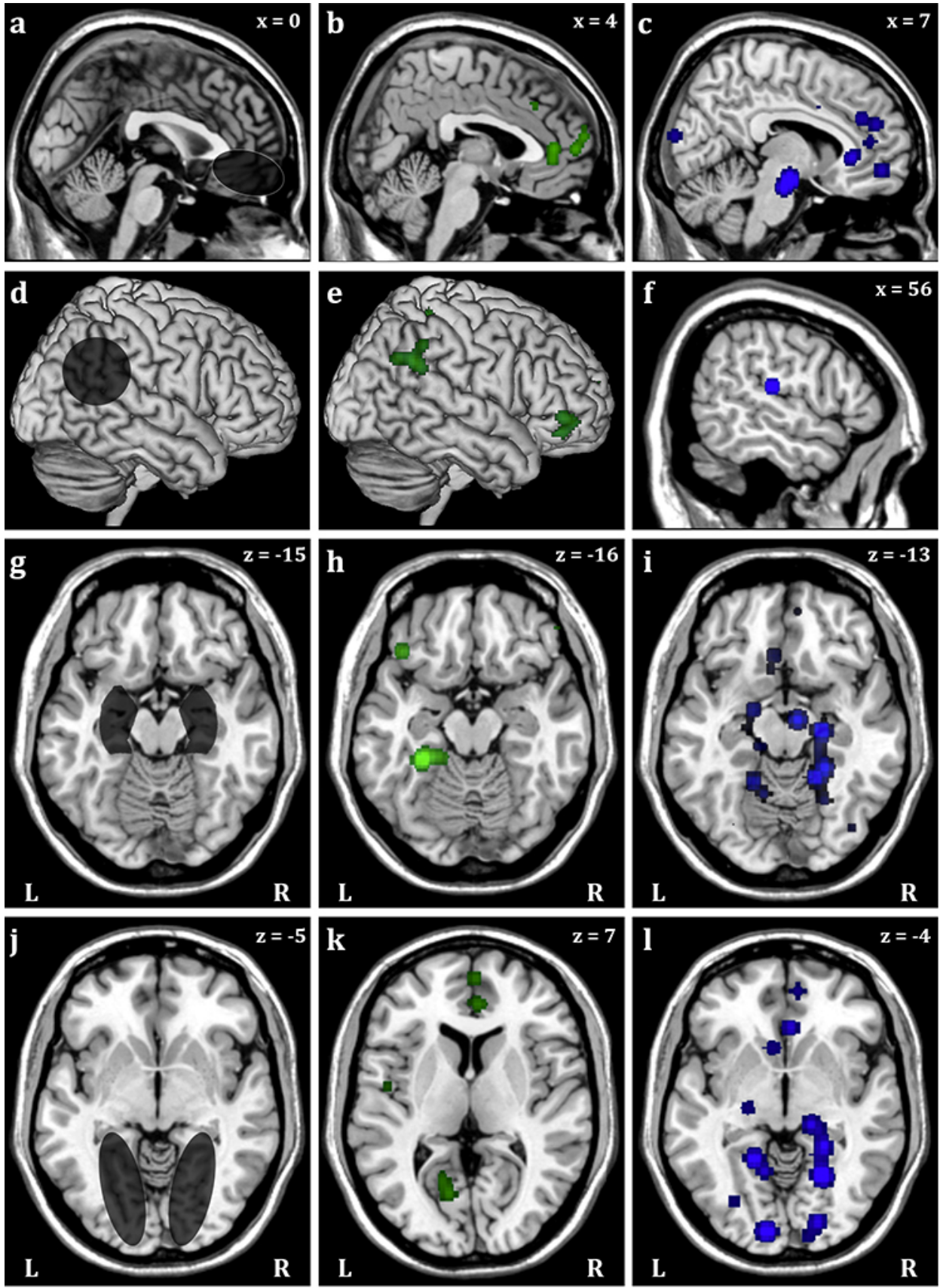

*Figure 2.* Neuropsychological lesion studies and meta-analysis of fMRI investigations converge on four brain structures that may be critical to self-generated thought.

The left column shows brain templates with dark areas indicating regions critical to nighttime self-generated thought (dreaming) as determined by overlapping lesion sites based on CT scans of neurological patients – except panel g, which shows the average site of medial temporal lobe excisions during surgery for intractable epilepsy. In the case of lesions to medial prefrontal cortex (panel a) there is evidence for severe reduction in waking self-generated thought (daydreaming/fantasy) as well. The middle column shows meta-analytic brain activations associated with waking daydreaming/mind-wandering. The right column shows meta-analytic brain activations associated with REM sleep, which is nearly always accompanied by dreaming ($r = .8$; Hobson et al., 2000). Note that the activation cluster in panel f is somewhat anterior to the temporoparietal junction, approximately in Brodmann area 40. All three approaches converge on four areas: the medial prefrontal cortex (panels a-c); the temporoparietal junction/inferior parietal lobule (panels d-f); the medial temporal lobe (panels g-i); and the medial occipital lobes/lingual gyrus (panels j-l). Panels a and d based on the work of Solms (2000a); panel j based on Solms (1997). Panels b, e, and h based on data from Fox et al. (2015). Panel g based on average medial temporal lobe excisions for severe epilepsy patients as reviewed by Mathern and Miller (2013). Panels c, f, and i based on data from Fox et al. (2013) and Domhoff & Fox (2015). X and Z values represent left-right and vertical coordinates, respectively, in Montreal Neurological Institute stereotactic space. Figure expanded and modified from Domhoff & Fox (2015), with permission.

*Figure 3.* Preferential involvement of medial temporal lobe structures and temporopolar cortex in electrophysiological stimulations (or spontaneous discharges) eliciting memories, thoughts, or hallucinatory, dream-like experiences.

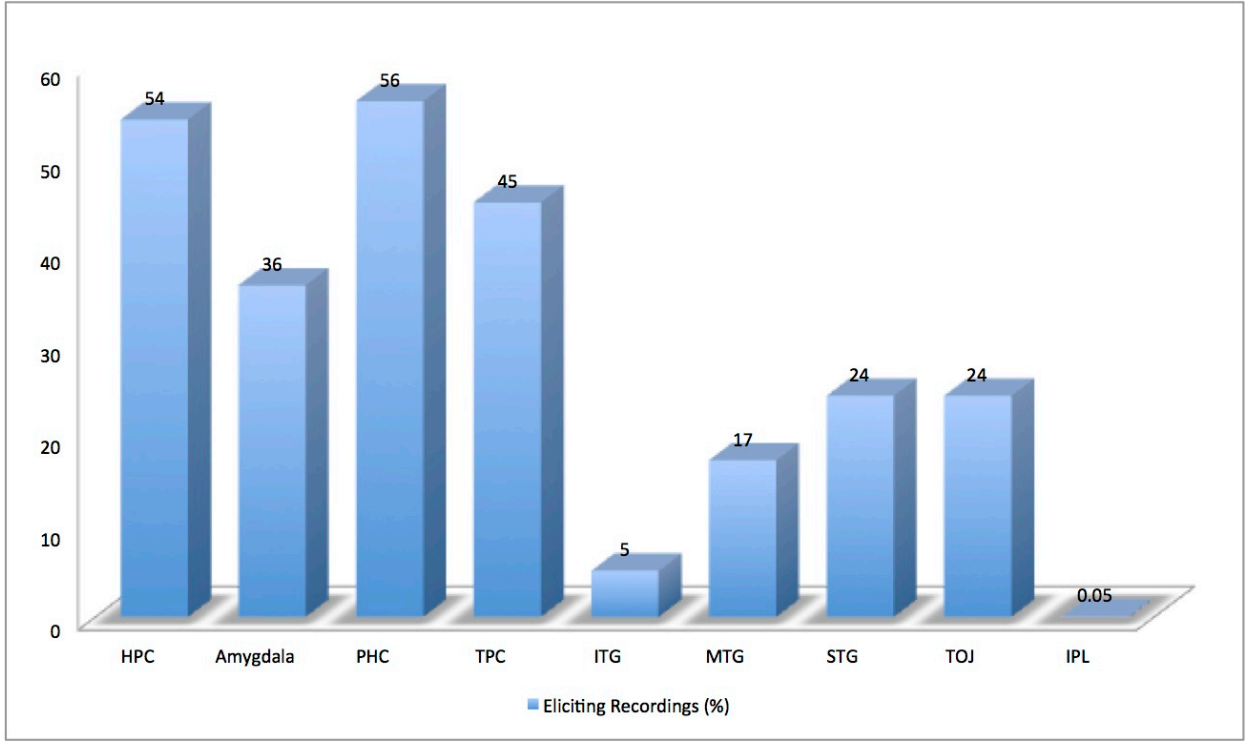

Percentage of stimulations or spontaneous discharges that elicited a first-person experience of memories, thoughts, or hallucinatory, dream-like experiences, based on more than 100 independent investigations. Not shown are data for hundreds of other stimulations throughout the brain, for which no such thought- or dream-like experiences have ever been reported. Only brain areas with ≥10 stimulations or discharges reported in the literature are visualized. Drawn from data in our Table 2, based on data in Supplementary Table 1 of the comprehensive review of Selimbeyoglu & Parvizi (2010). HPC: hippocampus; IPL: inferior parietal lobule; ITG: inferior temporal gyrus; MTG: middle temporal gyrus; PHC: parahippocampal cortex; STG: superior temporal gyrus; TOJ: temporo-occipital junction; TPC: temporopolar cortex.